\documentstyle[12pt]{article}
\def\be{\begin{equation}}
\def\ee{\end{equation}}
\def\bea{\begin{eqnarray}}
\def\eea{\end{eqnarray}}
\def\LM{\Lambda}

\def\pa{\partial}
\def\cL{{\cal L}}
\begin{document}
\renewcommand{\thefootnote}{\fnsymbol{footnote}}
\begin{titlepage}
\begin{flushright}
hep-th/9603073 \\
UTHEP-330 \\
March, 1996
\end{flushright}
\begin{center}
{\Large \bf Picard-Fuchs Equations and Prepotentials \\ in $N=2$ 
Supersymmetric QCD\footnote{
To appear in the proceedings of the Workshop \lq\lq
Frontiers in Quantum Field Theory'', Osaka, Japan,
December 14-17, 1995}}

\bigskip

Katsushi Ito and Sung-Kil Yang 

\medskip

{\it Institute of Physics, University of Tsukuba\\ Ibaraki 305, Japan}
\end{center}

\bigskip

\bigskip

\begin{abstract}
The Picard-Fuchs equations for $N=2$ supersymmetric $SU(N_{c})$ 
Yang-Mills theories with massless hypermultiplets are
obtained for $N_{c}=2$ and $3$. 
For $SU(2)$ we derive the non-linear differential equations for the
prepotentials and calculate full non-perturbative corrections to the
effective gauge coupling constant in the weak and strong coupling regions.
\end{abstract}
\end{titlepage}
\renewcommand{\thefootnote}{\arabic{footnote}}
\setcounter{footnote}{0}

\section{Introduction}
The work of Seiberg and Witten \cite{sewi,sewi2} on the exact solution 
to the low-energy effective action of $N=2$ supersymmetric Yang-Mills theory 
has afforded not only a renewed insight into the charge confinement 
\cite{sewi2} and the chiral symmetry breaking \cite{sewi2}, 
but also a marvelous insight into topological 
invariants \cite{wi} and conformal field theories in four 
dimensions \cite{scft}. The key ingredients in obtaining these exact results
are duality and the appearance of massless monopoles/dyons in the strong
coupling regions of the theory. 
In the weak coupling region, the exact solution enables us to
determine full non-peturbative instanton corrections to the 
effective coupling constant, whose evaluation is otherwise quite cumbersome
in the standard framework of quantum field theory.

The low-energy effective action is described in terms of a prepotential
which is a single holomorphic function of superfields of  $N=2$ 
$U(1)$ vector multiplets. 
The exact solution for the prepotential may be 
characterized by the period integrals of 
the special type of one-form on a hyper-elliptic curve.
The curves associated with a variety of $N=2$ 
supersymmetric Yang-Mills theories and QCD 
have been studied extensively \cite{KlLeThYa}$-$\cite{itsa}.
The moduli space of the curve contains singularities, at which 
some solitons become massless. 
In order to investigate the strong coupling physics, one needs to 
evaluate the period integrals near the singular locus.
An efficient approach to study this problem is to use the Picard-Fuchs 
equation, the differential 
equation which the periods obey \cite{Cer,KlLeTh,Ma,ItYa}.

In the present article, we shall investigate the quantum moduli space
of $N=2$ supersymmetric Yang-Mills theories with massless hypermultiplets
and gauge groups $SU(2)$ and $SU(3)$.
We shall derive the Picard-Fuchs equation for the scalar part of $N=2$ 
$U(1)$ vector multiplets and their duals.
By solving the non-linear differential equation obeyed by the prepotential
for $G=SU(2)$, we will explicitly evaluate
the non-perturvative contributions in the prepotential 
in both weak and strong coupling regions.

\section{$N=2$ Supersymmetric Yang-Mills Theory with Massless Hypermultiplets}
We begin with reviewing some basic properties of the low-energy 
effective action of the $N=2$ supersymmetric $G=SU(N_{c})$ QCD
\cite{sewi2,arplsh}. 
In $N=1$ superfield formulation, 
the theory contains chiral 
multiplets $\Phi^{a}$ and  chiral field strength $W^{a}$ 
($a=1,\ldots, {\rm dim}G$) both in the adjoint representation of $G$, 
and chiral superfields $Q^{i}$ in the $N_{c}$ and 
$\tilde{Q}^{i}$ ($i=1,\cdots, N_{f}$)
in the $\bar{N}_{c}$ representation of the gauge group.
The superpotential is given by 
${\cal W}=\tilde{Q}T^{a}\Phi^{a}Q+M^{i}_{j}\tilde{Q}_{i} Q^{j}$
where $T^{a}$ is the generator of $G$ and $M^{i}_{j}$ is the mass matrix.
Along the flat direction the scalar fields $\phi$ of $\Phi$ get
vacuum expectation values, which break the gauge group to the Cartan subgroup 
$U(1)^{r}$ where $r=N_{c}-1$ is the rank of $G=SU(N_{c})$. 
When the squark fields do not have vacuum expectation values, the low-energy
effective theory is in 
the Coulomb branch and contains $r$ $U(1)$ vector multiplets 
$(A^{i}, W_{\alpha}^{i})$ $(i=1,\cdots,r)$, where $A^{i}$ are $N=1$
chiral superfields and $W_{\alpha}^{i}$ are $N=1$ vector superfields. 
The quantum moduli space may be characterized by 
the low-energy effective Lagrangian ${\cal L}$ 
with the prepotential ${\cal F}(A)$
\be
{\cal L}={1\over 4\pi} {\rm Im }
\left(\int d^2\theta d^2\bar{\theta}A_{D i}\bar{A}^{i}+{1\over2}
\int d^2\theta \tau^{i j}W_{\alpha}^{i} W^{j \alpha}\right), 
\ee
where $A_{D i}={\pa {\cal F}\over \pa A^{i}}$ is a field dual to $A^{i}$ 
and 
$\tau^{i j}={\pa^{2}{\cal F}\over \pa A^{i} \pa A^{j}}$ the effective
coupling constants.
We denote the scalar component of $A^{i}$, $A_{D i}$ by $a^{i}$, $a_{D i}$.
The pairs $(a_{D i},a_{i})$ are the $Sp(2r,{\bf Z})$ section over the 
space of gauge invariant parameters $s_{i}$ $(i=2,\cdots, N_{c})$
defined by ${\rm det}(x-\phi)=x^{N_{c}}-\sum_{i=2}^{N_{c}} s_{i} x^{N_{c}-i}$.
The quantum moduli space of the Coulomb branch is parametrized by the 
gauge invariants $s_{i}$ and the eigenvalues $m_{1},\ldots, m_{N_{f}}$ 
of the mass matrix.

The sections $(a_{D i},a_{i})$ are obtained as
the period integrals of a meromorphic differential
$\lambda$ over the hyper-elliptic curve ${\cal C}$ \cite{haoz}:
\bea
y^{2}&=&C(x)^2-G(x), \nonumber \\
C(x)&=& x^{N_{c}}-\sum_{i=2}^{N_{c}} s_{i} x^{N_{c}-i}
       +{\Lambda^{2N_{c}-N_{f}}\over 4} \sum_{i=0}^{N_{f}-N_{c}}
 t_{i}(m) x^{N_{f}-N_{c}-i},
\nonumber \\
G(x)&=& \Lambda^{2N_{c}-N_{f}}\prod_{i=1}^{N_{f}}(x+m_{i}),
\eea
where $\Lambda$ stands for the QCD scale parameter and $t_{i}(m)$ is defined by
$\prod_{i=1}^{N_{f}}(x+m_{i})=\sum_{i=0}^{N_{f}} t_{i}(m) x^{N_{f}-i}$ with
$t_{0}(m)=1$.
The terms proportional to $\Lambda^{2N_{c}-N_{f}}$ in $C(x)$ are absent 
in the case of $N_{f}<N_{c}$. Notice that for $N_c=2$, $N_f=2$ theory
$C(x)$ should read
\be
C(x)=x^2-s_2+{\LM^2\over 8}.
\ee

The meromorphic differential $\lambda$ is given by
\be
\lambda={x\over 2\pi i} d\log {C-y\over C+y}.
\ee
This differential satisfies the relations
\be
{\pa \lambda \over \pa s_{i}}=-\omega_{N_{c}+1-i}+d(*),
\ee
where $\omega_{i}={x^{i-1}dx \over y}$ is the basis of the holomorphic 
one-forms of the hyper-elliptic curve ${\cal C}$.
Choosing the basis of homology cycles $(A_{i},B_{i})$ on ${\cal C}$ with
the canonical intersection form $A_{i}\cdot B_{j}=\delta_{i j}$, 
$A_{i}\cdot A_{j}=B_{i}\cdot B_{j}=0$, one can express 
the pairs $(a_{D i},a_{i})$ as
\bea
\alpha_{i}\cdot a_{D} &= & \int_{B_{i}}\lambda, \nonumber \\
(\lambda_{i}-\lambda_{i-1})\cdot a &=& \int_{A_{i}}\lambda,  \hskip7mm
i=1,\cdots,r,
\eea
where $\alpha_{i}$ are the simple roots of $G$ and $\lambda_{i}$ 
the fundamental weights of $G$ ($\lambda_{0}\equiv0$).

The BPS mass formula $M\geq |q_{i}a_{i}+h_{i}a_{D i}|$
shows that the massless soliton may be characterized by the vanishing cycle
$q_{i}A_{i}+h_{i}B_{i}\sim 0$.
Therefore the singularities of the quantum moduli space are determined by the 
discriminant $\Delta$ of the curve.
Let us enumerate explicit forms of the discriminants $\Delta^{N_{c}}_{N_{f}}$ 
for the $SU(N_c)$ theories ($N_c =2,\, 3$) with $N_{f}(<2 N_{c})$ matter
hypermultiplets. In the following 
$\LM_{N_{f}}$ denotes the scale parameter corresponding to 
the $N_{f}$-flavor theory.
\begin{itemize}
\item $N_{c}=2$ ($u\equiv s_{2}$)
\bea
\Delta^{2}_{0}&=& 256 \LM_{0}^{8} (u-\LM_{0}^2)(u+\LM_{0}^2)
\nonumber \\
\Delta^{2}_{1}&=& -\LM_{1}^{6} (27 \LM_{1}^{6}+256 u^3) \nonumber \\
\Delta^{2}_{2}&=& \LM_{2}^4 (8 u-\LM_{2}^2)^2 (8 u+\LM_{2}^2)^2/16
\nonumber \\
\Delta^{2}_{3}&=& \LM_{3}^{2} (\LM_{3}^{2}-256 u) u^4 \nonumber \\
\eea
\item $N_{c}=3$ ($u\equiv s_{2}$, $v\equiv s_{3}$)
\bea
\Delta^{3}_{0}&=& 64 \LM_{0}^{18}
       (-27 (v-\LM_{0}^3)^2+4 u^3)
                    (-4 u^3+27 (v+\LM_{0}^3)^2)
\nonumber \\
\Delta^{3}_{1}&=&\LM_{1}^{15}
 (-3125 \LM_{1}^{15}+ 256 \LM_{1}^{5}  u^{5}
  + 22500 \LM_{1}^{10}   u v - 1024 u^{6}  v \nonumber \\
& & -
 43200 \LM_{1}^{5}  u^2  v^2  + 13824 u^3  v^3  - 46656 v^5 ) \nonumber \\
\Delta^{3}_{2}&=&  64 \LM_{2}^{12} v^2 (-4 (\LM_{2}^2-u)^3 - 27 v^2) 
(-4 (\LM_{2}^2+u)^3+ 27 v^2)
\nonumber \\
\Delta^{3}_{3}&=& \LM_{3}^9  (-\LM_{3}^3  + 4 v)^3
   (729 \LM_{3}^6  u^3  - 256 u^6  - 3888 \LM_{3}^3  u^3  v \nonumber \\
  & & + 3456 u^3  v^2  +
      2916 \LM_{3}^3  v^3  - 11664 v^4 )) / 16 \nonumber \\
\Delta^{3}_{4}&=& 4 \LM_{4}^6 v^4 (\LM_{4}^4 u - 8 \LM_{4}^2 u^2 +
 16 u^3 + 2 \LM_{4}^3 v - 72 \LM_{4} u v - 108 v^2)\nonumber  \\
& &    (-\LM_{4}^4 u + 8 \LM_{4}^2 u^2 - 16 u^3 + 2 \LM_{4}^3 v
- 72 \LM_{4} u v + 108 v^2) \nonumber \\
\Delta^{3}_{5}&=& 4 \LM_{5}^3  v^5  (\LM_{5}^2  u^5  - 256 u^6  +
2 \LM_{5}^3  u^3  v - 528 \LM_{5} u^4  v +
 \LM_{5}^4  u v^2 \nonumber \\
& &  -
300 \LM_{5}^2  u^2  v^2  + 3456 u^3  v^2  + 4 \LM_{5}^3  v^3  -
4212 \LM_{5} u v^3  - 11664 v^4 ). \nonumber \\
\eea
\end{itemize}
Note that for the gauge group $SU(2)$ the zeros of the discriminant realize 
the discrete
${\bf Z}_{4-N_{f}}$ symmetry which arises from the $U(1)_{\cal R}$ 
anomaly and the parity transformation in the flavor symmetry $O(2 N_{f})$.
Order of the zeros corresponds to the number
of massless hypermultiplets appearing at the singularities. 
Massless monopoles obey the spinor representation of the flavor 
symmetry $SO(2 N_{f})$. 
For $SU(3)$ symmetries are less obvious  particularly for odd flavors.

\section{The Picard-Fuchs equations}
Let us study the Picard-Fuchs equations for 
$(a_{D i},a_{i})$ and examine their solutions.
We first discuss the $SU(2)$ case in detail for simplicity. 
One can change the curve of fourth order to the third one by 
using the M\"obius transformation. 
The curves are given by
\be
y^2=x^2(x-u)+{1 \over 4} \Lambda_0^4 \, x
\ee
for $N_f=0$ and
\be
y^2=x^2(x-u)-{1 \over 64} \Lambda_{N_f}^{2(4-N_f)} (x-u)^{N_f-1}
\ee
for $N_f=1,\, 2,\, 3$.
The meromorphic differential $\lambda$ for this curve
is given by 
\be
\lambda ={\sqrt{2} \over 8\pi}\, {2u-(4-N_f)x \over y}\, dx
\ee
which satisfies
\be
{d\lambda \over du}={\sqrt{2} \over 8\pi} {d x \over y}+d(*).
\ee
It is not difficult to see that the period $\Pi=\oint \lambda$ satisfies the 
differential equation 
\be
p(u){d^2 \Pi \over du^2}+ \Pi =0,
\label{pfsu2}
\ee
where \cite{Cer,KlLeTh,ItYa}
\be
p(u)=\left\{ \begin{array}{ll}
\displaystyle{4 (u^2-\LM_{0}^4)} & \qquad N_{f}=0 \\
\displaystyle{4u^2+{27 \LM_1^6\over 64u}} & \qquad N_{f}=1 \\
\displaystyle{\ 4\Big(u^2-{\LM_2^4\over 64}\Big)} & \qquad N_{f}=2 \\
\displaystyle{u\Big(4u-{\LM_3^2\over 64}\Big)} & \qquad N_{f}=3.
\end{array}
\right.
\ee
Note that the regular singularities of (\ref{pfsu2}) correspond to the
zeros of the discriminant. Introducing a  new variable
\be
z=\left\{\begin{array}{ll}
\displaystyle{\left({u\over \LM_{0}^2}\right)^2} & \qquad N_{f}=0 \\
\ \displaystyle{-{256\over 27} \left({u\over \LM_{1}^2}\right)^3}
 & \qquad N_{f}=1 \\
\displaystyle{64 \left({u\over \LM_{2}^2}\right)^2 }& \qquad N_{f}=2 \\
\displaystyle{256 \left({u\over \LM_{3}^2}\right)} & \qquad N_{f}=3, \\
\end{array}
 \right.
\ee
it is shown that the Picard-Fuchs equation (\ref{pfsu2}) turns out to be the
Gauss's hypergeometric system
\be
z(1-z) {d^{2} \Pi \over d z^{2}}
+(\gamma-(\alpha+\beta+1)z){d \Pi \over d z}-\alpha\beta \Pi=0,
\label{hyperg}
\ee
where
\be
\alpha=\beta={-1\over 2(4-N_{f})}, \hskip10mm  
\gamma={3-N_{f} \over 4-N_{f}}.
\ee
Using the fundamental solutions of (\ref{hyperg}), we may evaluate 
the power series solutions for $(a_{D},a)$ near the singularities, 
which has been  explicitly done in our previous work \cite{ItYa}.

Next we write down the Picard-Fuchs equations for $SU(3)$ theory 
with flavors $N_{f}\leq 5$.
The periods $\Pi=\oint \lambda$ now depend on two parameters $u$ and $v$.
The Picard-Fuchs equations take the form
\be
{\cal L}_{1}\Pi=0, \quad {\cal L}_{2}\Pi=0,
\ee
where $\cL_{1}$ and $\cL_{2}$ are given by
\begin{itemize}
\item $N_{f}=0$
\bea
\cL_{1}&=& {1\over 3} (-27\LM_{0}^{6}+4 u^3+27 v^2)\pa_{v}^{2}
          +12 u v \pa_{u}\pa_{v}
          +3 v \pa_{v}+1 \nonumber \\
\cL_{2}&=& {1\over u } (-27\LM_{0}^{6}+4 u^3+27 v^2)\pa_{u}^{2}
          +12 u v \pa_{u}\pa_{v}
          +3 v \pa_{v}+1
\eea
\item $N_{f}=1$
\bea
\cL_{1}&=& {-25\LM_{1}^{5}u^2+84 u^3 v +405 v^3 \over 45 v}\pa_{v}^{2}
          +{64 u^4-1125 \LM_{1}^{5} v+2160 u v^{2}\over 180 v}
 \pa_{u}\pa_{v} \nonumber \\
       & &+ {16 u^3+135 v^2 \over 45 v} \pa_{v}+1  \nonumber \\
\cL_{2}&=& {4 (-25\LM_{1}^{5}u^2+84 u^3 v +405 v^3)\over
            -25 \LM_{1}^5 +84 u v} \pa_{u}^{2} \nonumber \\
& &
+{(-25 \LM_{1}^5 +36 u v) (-25 \LM_{1}^5+96 u v)\over -25 \LM_{1}^{5} +84 u v}
 \pa_{u}\pa_{v} \nonumber \\
 & &    +  {3 v (-25 \LM_{1}^5 +36 u v)\over -25 \LM_{1}^5 +84 u v} \pa_{v}
          +1
\eea
\item  $N_{f}=2$
\bea
\cL_{1}&=& {-8\LM_{2}^{4}u+8 u^3  +27 v^2\over 3}\pa_{v}^{2}
+{4 (2 \LM_{2}^8 -4 \LM_{2}^4 u^2 +2 u^4+27 u v^2)\over 9 v}
 \pa_{u}\pa_{v} \nonumber \\
  & &     + { -8 \LM_{2}^4 u +8 u^3 +27 v^2 \over 9 v} \pa_{v}+1  \nonumber \\
\cL_{2}&=& {-8\LM_{2}^{4}u+8 u^3  +27 v^2\over 2 u} \pa_{u}^{2}
          + {3v (\LM_{2}^4+3 u^2)\over u} \pa_{u}\pa_{v}     +  1
\eea
\item  $N_{f}=3$
\bea
\cL_{1}&=& {( 4 v-\LM_{3}^3) (4 u^3+9 v^2)\over 4 v} \pa_{v}^{2}
+{u (64 u^3-81 \LM_{3}^3 v +432  v^2) \over 36 v}  \pa_{u}\pa_{v} \nonumber \\
       & &     + { 16 u^3 +27 v^2 \over 9 v} \pa_{v}+1  \nonumber \\
\cL_{2}&=& {4 u^3+9 v^2\over u} \pa_{u}^{2}
          +{u (-9 \LM_{3}^3 +32  v)\over 4} \pa_{u}\pa_{v}
          - v  \pa_{v}
                +1
\eea

\item  $N_{f}=4$
\bea
\cL_{1}&=& { -\LM_{4}^4 u -56 \LM_{4}^2 u^2+240 u^3+324 v^2\over 36} \pa_{v}^{2}
+ {-8 \LM_{4}^2 u^2 +32 u^3+27 v^2 \over 9 v} \pa_{v} \nonumber \\
& &
 +{2 \LM_{4}^4 u^2 -16 \LM_{4}^2 u^3 +32 u^4
  -9 \LM_{4}^2 v^2  +108 u v^2\over 9 v} \pa_{u}\pa_{v}
         +1  \nonumber \\
\cL_{2}&=&
{ -\LM_{4}^4 u -56 \LM_{4}^2 u^2+240 u^3+324 v^2 \over \LM_{4}^2+60 u}
 \pa_{u}^{2} \nonumber \\
& & +{v(\LM_{4}^2+12 u) (-\LM_{4}^2+36 u) \over \LM_{4}^2+60 u}\pa_{u}\pa_{v}
     +{3 v (\LM_{4}^{2}-36 u)\over \LM_{4}^2+60 u} \pa_{v}+1 \nonumber \\
\eea

\item
 $N_{f}=5$
\bea
\cL_{1}&=& {v (-2\LM_{5}^2 u^2 +528 u^3+57\LM_{5} u v +324 v^2)\over
 5 \LM_{5} u + 36 v}\pa_{v}^{2} \nonumber \\
& &     +{ -5 \LM_{5}^2 u^3 +1280 u^4 -\LM_{5}^3 u v +332 \LM_{5} u^2 v
-4 \LM_{5}^2 v^2+1728 u v^2 \over 4 (5 \LM_{5} u
 + 36 v)} \pa_{u}\pa_{v} \nonumber \\
 & & +
{-5 \LM_{5}^2 u^2 +1280 u^3+88 \LM_{5} u v 
+432 v^2 \over 4(5 \LM_{5} u + 36 v)} \pa_{v}+
1  \nonumber \\
\cL_{2}&=& 
{-2\LM_{5}^2 u^2 +528 u^3+57\LM_{5} u v +324 v^2\over 132 u}
 \pa_{u}^2 \nonumber \\
& & +{-\LM_{5} u^3-5 \LM_{5}^2 u v+864 u^2 v-36 \LM_{5} v^{2}
\over 132 u}\pa_{u}\pa_{v}
 -{\LM_{5} u +324 v\over 132} \pa_{v}+1. \nonumber \\
\eea
\end{itemize}
The Picard-Fuchs equations  for $N_{c}=3$, $N_{f}=0$ theory
have been investigated in detail by Klemm, Lerche and Theisen \cite{KlLeTh}.
They show that the system reduces to the Appell's hypergeometric system of 
type $F_{4}$. Although other cases are not classified as the Appell's system, 
the solutions of
these equations may be obtained in the form of the power series expansion 
in $u$ and $v$.

\section{Calculation of the Prepotentials}
Solving the system of the Picard-Fuchs equations, we evaluate the 
non-perturbative  corrections to the prepotentials, which can be 
regarded as the multi-instanton contributions in the weak coupling 
region and the threshold corrections in the strong coupling regions.
In the previous work \cite{ItYa}, we have calculated some 
non-trivial  corrections by using the solutions of the hypergeometric system. 
In this article, we shall instead take more direct approach by solving the 
non-linear differential equation for the prepotential in $SU(2)$ theory. 
The present approach was proposed originally by Matone for $SU(2)$ pure Yang-
Mills theory \cite{Ma}.

We start with the Picard-Fuchs equation (\ref{pfsu2}). 
Notice that the equation carries no first derivative term. 
This implies that the Wronskian of the system
\be
W=\left|
\begin{array}{cc}
a & a_{D} \\
\pa_{u}a & \pa_{u}a_{D}
\end{array} \right|
\ee
is independent of $u$; $\pa_{u}W=0$. When we integrate this with respect to
$u$ there appears a constant, the explicit value of which 
is evaluated at any singularity in the moduli space. One finds that
\be
W={i b \over 4\pi},
\label{wrons}
\ee
where $b=4-N_{f}$ is the coefficient of the beta-function of $SU(2)$ $N=2$ 
QCD.  
In the present case, we can further simplify the equation (\ref{wrons}) since
the Wronskian is written in the form of the total derivative
\be
W=\pa_{u} \left( a {\pa {\cal F} \over \pa a} -2 {\cal F}\right).
\label{totald}
\ee
Thus we get the following relation between the prepotential and the 
gauge invariant parameter $u$
\be
a {\pa {\cal F} \over \pa a} -2 {\cal F}={i b \over 4\pi}u + {\rm const.}
\label{scaling}
\ee
The constant term may be absorbed in the definition of ${\cal F}$ by shifts,
which does not affect the form of the low-energy effective action.
This identity has been generalized for any $N=2$ Yang-Mills theories with 
or without matter hypermultiplets \cite{sothya,egya} 
and is important in view of 
its relation to the soliton theory \cite{whto}.

The relation (\ref{scaling}) allows us to express $u$ in terms of $a$
\be
u={\cal G}(a)\equiv {4\pi \over i b}
\left(a {\pa {\cal F} \over \pa a} -2 {\cal F}\right).
\ee
The Picard-Fuchs equation becomes
\be
-p({\cal G}(a)){d^2 {\cal G} \over d a^2}+
a \left({d {\cal G}\over d a}\right)^3=0.
\label{nonl}
\ee
This non-linear equation is utilized to  determine the instanton coefficients 
recursively.
Near the singularity at $u=\infty$, the prepotential takes the form
\be
{\cal F}(a)=
\left\{\begin{array}{ll}
{i a^{2}\over 2\pi}
\left\{ 2 \log {a^{2}\over \Lambda_{0}^{2}}
+\sum_{k=0}^{\infty} {\cal F}_{k} (0)
\left( {\Lambda_{0}\over a} \right)^{4k}\right\}
 & N_{f}=0 \\
{i a^{2}\over 4\pi}
\left\{ (4-N_{f}) \ln {a^{2}\over \Lambda_{N_{f}}^{2}}
+\sum_{k=0}^{\infty} {\cal F}_{k} (N_{f})
\left( {\Lambda_{N_{f}}^{2} \over a^{2}} \right)^{k(4-N_{f})}\right\}
& N_{f}=1,2,3.
\end{array}\right.
\ee
The first four coefficients are listed in the table
\begin{center}
\begin{tabular}{|c|c|c|c|c|}
\hline
$N_{f}$ & ${\cal F}_{1}$ & ${\cal F}_{2}$ & ${\cal F}_{3}$ & ${\cal F}_{4}$ \\
\hline
        &                              & & & \\
0       &  $-{1\over 2^5}$              & $-{5\over 2^{14}}$    &
        $-{3\over 2^{18}}$ & $-{1469\over 2^{18}}$  \\
        &                              & & & \\
1       &  ${3\over 2^{12}}$             & -${3\cdot 17 \over 2^{288}}$  &
           ${5\cdot 7 \cdot 11\over 3 \cdot 2^{38}}$ &
           ${3^2 \cdot  5 \cdot 7\cdot 9679 \over 2^{59}}$  \\
        &                              & & & \\
2       &  $-{1\over 2^{11}}$
        &  $ -{5\over 2^{26}}$
        &  $-{3\over 2^{36}}$
        &  $ -{13\cdot 113 \over 2^{55}}$ \\
        &                              & & & \\
3       &  $-{1\over 2^{5}}$
        &  $-{1\over 2^{24}}$
        &  $0$
       &   $-{5\over 2^{51}}$\\
\hline
\end{tabular}
\end{center}
and summarized by the formulas ($N_{f}=1,2,3$)
\footnote{We have corrected an error of  ${\cal F}_{2}(N_{f})$ in the 
previous work \cite{ItYa}.}:
\bea
{\cal F}_{1} (N_{f})&=&-{1  \over 2^{b} c(N_{f})}
                         {1\over 2\, b^2}, \nonumber \\
{\cal F}_{2} (N_{f})&=&-{1\over 2^{2b} c(N_{f})^{2}}
                       {5-8b+4b^{2} \over
                         8^{2}\, b^{4}}, \nonumber \\
{\cal F}_{3} (N_{f})&=&{1\over 2^{3b}c(N_{f})^{3}}
                      {   (1-b)(1-2b)
                         (-23+30b-16b^{2}) \over
                       1728\, b^{6}}, \nonumber \\
{\cal F}_{4} (N_{f})&=&{1\over 2^{4b} c(N_{f})^{4}}
{(1 - 2b )(-677 + 3910b - 8124b^2 +
      8456b^3 - 4672b^4 + 1152b^5) \over
294912\, b^8}, \nonumber \\
\label{weakcorre}
\eea
where 
$c(N_{f})=\eta_{N_{f}} 2^{8} (4-N_{f})^{N_{f}-4}$ ($N_{f}=1,2,3$) and
$\eta_{1}=-1,\, \eta_{2}=\eta_{3}=1$.
The $N_{f}=0$ result \cite{KlLeTh,Ma} may be 
obtained by replacing $c(N_{f})$ with 1 in the $N_{f}=2$ case.
The results (\ref{weakcorre}) indeed agree with those obtained previously
\cite{ItYa}.

Similarly we may solve the non-liner differential equation 
in the strong coupling region where monopoles become massless.
In this case the dual field $a_{D}$ is a good coordinate around the 
singularity $u=u_{0}$ where
\be
u_{0}=\left\{\begin{array}{ll}
\Lambda_{0}^2 & \qquad N_{f}=0 \\
\displaystyle{-3\cdot 2^{-{8\over 3}}\Lambda_{1}^2 }
 & \qquad N_{f}=1 \\
\displaystyle{\Lambda_{2}^2/8} & \qquad N_{f}=2 \\
0 & \qquad N_{f}=3. \\
\end{array}
 \right.
\ee
Introduce the dual prepotential ${\cal F}_{D}(a_{D})$ satisfying
$a={d {\cal F}_{D}\over d a_{D}}$. Then, instead of (\ref{totald}), the
Wronskian is expressed in terms of ${\cal F}_{D}$. Hence
around $u=u_{0}$ we get
\be
u=u_{0}+{4\pi\over b} \left( 
2  {\cal F}_{D} -a_{D}{d  {\cal F}_{D} \over d a_{D}} \right).
\ee
Similarly to the weak-coupling computation we now obtain the non-linear
differential equation which makes it possible to calculate ${\cal F}_{D}$
recursively. Using the expansion
\be
 {\cal F}_{D}={i \over 8\pi} a_{D}^2\left\{ k \log 
\left({a_{D}^2\over \Lambda_{N_{f}}^2}\right)
+\sum_{n\geq-1} {\cal F}_{D n}(N_{f}) 
\left({a_{D}\over \Lambda_{N_{f}}}\right)^n \right\}
\ee
with $k=1$ ($N_{f}=0$) and $k=2^{N_{f}-1}$ ($N_{f}=1,2,3$), 
we find the coefficients as follows
\bea
{\cal F}_{D\ -1}(N_{f})&=& 16 k b^2 \tilde{c}(N_{f}), \nonumber \\
{\cal F}_{D \ 1}(N_{f})&=&{2\cdot 2 k
 \over 3 \tilde{c}(N_{f})}
                         {(1-2b)(-5+2b)\over 16 b^{2}}, \nonumber \\
{\cal F}_{D \ 2}(N_{f})&=&{-2 k\over 2 \tilde{c}(N_{f})^{2}}
{(1-2b)(-61+190b-92 b^{2}+8b^{3})
\over 1152 b^{4}}, \nonumber \\
{\cal F}_{D \ 3}(N_{f})&=&{2\cdot 2k\over 5 \tilde{c}(N_{f})^{3}}
{(1-2b)(1379-8162b+14596 b^{2}-7288b^{3}+960b^{4})
\over 110592 b^{6}}, \nonumber \\
\eea
where
$\tilde{c}(1)=-3^{-1/2}\cdot 2^{-17/6}$,$\tilde{c}(2)=-i 2^{-9/2}$ and
$\tilde{c}(3)=2^{-13/2}$.
The $N_{f}=0$ result is obtained by putting $b=2$ and 
$\tilde{c}(0)=-i/4$.
Here we have corrected some misprints in the previous work \cite{ItYa}.

For $SU(3)$ theory the prepotential ${\cal F}$ obeys the basic identity
analogous to ({\ref{scaling}) \cite{sothya,egya}. 
In contrast to the $SU(2)$ case, combining
this with the $SU(3)$ Picard-Fuchs equations does not yield a simple
recursion relation. It is not clear at present if the existence of such
recursion relation is peculiar to $SU(2)$ theory. It will be interesting
if one could find an additional constraint equation for ${\cal F}$ which
eventually leads to the direct evaluation of instanton corrections.

\vskip5mm
\section*{Acknowledgments}
The work of K.I. is supported in part by University of Tsukuba
Research Projects and the Grant-in-Aid for Scientific Research from
the Ministry of Education (No. 07210210).
The work of S.-K.Y. is supported in part by Grant-in-Aid for Scientific
Research on Priority Area 231 ``Infinite Analysis'',
the Ministry of Education, Science and Culture, Japan.


\begin{thebibliography}{99}
\bibitem{sewi} N.~Seiberg and E.~Witten, Nucl. Phys. {\bf B426} (1994) 19.
\bibitem{sewi2} N.~Seiberg and E.~Witten, Nucl. Phys. {\bf B431} (1994) 484.
\bibitem{wi}
E.~Witten, J. Math. Phys. {\bf 35} (1994) 5101; 
Math. Res. Lett. {\bf 1} (1994) 769.
\bibitem{scft}
P.C. Argyres and M.R. Douglas, Nucl. Phys. {\bf  B448} (1995) 93;

P.C. Argyres, M.R. Plesser, N. Seiberg and E. Witten, {\it New $N=2$
Superconformal Field Theories in Four Dimensions}, hep-th/9511154;

T.~Eguchi, K.~Hori, K.~Ito and S.-K.~Yang, 
{\it Study of $N=2$ Superconformal Field Theories in $4$
Dimensions}, hep-th/9603002.

\bibitem{KlLeThYa}
A. Klemm, W. Lerche, S. Theisen and S. Yankielowicz,
Phys. Lett. {\bf B344} (1995) 169; 

P.C. Argyres and A.E. Faraggi, Phys. Rev. Lett. {\bf 74} (1995) 3931;

\bibitem{dasu}
U.H.~Danielsson and B.~Sundborgm,   Phys. Lett. {\bf B358} (1995) 273.

\bibitem{dash}
M.R.~Douglas and S.H.~Shenker,   Nucl. Phys. {\bf B447} (1995) 271.

\bibitem{brla}
A.~Brandhuber and K.~Landsteiner,   Phys. Lett. {\bf B358} (1995) 73.

\bibitem{haoz}
A.~Hanany and Y.~Oz,   Nucl. Phys. {\bf B452} (1995) 283.

\bibitem{arplsh}
P.C.~Argyres, M.R.~Plesser and A.D.~Shapere,
Phys. Rev. Lett. {\bf 75} (1995) 1699.

\bibitem{Cer}
A. Ceresole, R. D'Auria and S. Ferrara, Phys. Lett. {\bf B339}
(1994) 71.

\bibitem{KlLeTh}
A. Klemm, W. Lerche and S. Theisen, {\it Nonperturbative
Effective Actions of $N=2$ Supersymmetric Gauge Theories}, hep-th/9505150.

\bibitem{Ma} M.~Matone,   Phys. Lett. {\bf B357} (1995) 342.

\bibitem{ItYa} K.~Ito and S.-K.~Yang,   Phys. Lett. {\bf B366} (1996) 165.

\bibitem{mine}
J.A.~Minahan and D. Nemeschansky, {\it Hyperelliptic Curves for
Supersymmetric Yang-Mills}, hep-th/9507032.

\bibitem{arsh}
P.C.~Argyres and A.D.~Shapere, 
{\it The Vacuum Structure Of $N=2$ Super-QCD With Classical Gauge
Groups}, hep-th/9509175.

\bibitem{ha}
A.~Hanany, {\it On The Quantum Moduli Space of Vacua
of $N=2$ Supersymmetric Gauge Theories}, hep-th/9509176.

\bibitem{dasu2}
U.H.~Danielsson and B.~Sundborg, {\it Exceptional Equivalences in $N=2$
Supersymmetric Yang-Mills Theory}, hep-th/9511180.



\bibitem{alar}
M. Alishahiha, F. Ardalan and F. Mansouri, {\it The Moduli Space of the
$N=2$ Supersymmetric $G_2$ Yang-Mills Theory}, hep-th/9512005.

\bibitem{itsa}
K.~Ito and N.~Sasakura, {\it One-Instanton Calculations in $N=2$
Supersymmetric $SU(N_c)$ Yang-Mills Theory}, hep-th/9602073.


\bibitem{sothya}
J.~Sonnenschein, S.~Theisen and S.~Yankielowicz, 
Phys. Lett. {\bf B367} (1996) 145.

\bibitem{egya}
T.~Eguchi and S.-K.~Yang, {\it Prepotentials  of $N=2$ Supersymmetric 
Gauge Theories and Soliton Equations}, hep-th/9510183.

\bibitem{whto}
A. Gorsky, I. Krichever, A. Marshakov, A. Mironov and A. Morozof,
Phys. Lett. {\bf B355} (1995) 466;

T. Nakatsu and K. Takasaki, {\it Whitham-Toda Hierarchy and $N=2$
Supersymmetric Yang-Mills Theory}, hep-th/9509162;

E. Martinec and N. Warner, Nucl. Phys. {\bf B459} (1996) 97; 
{\it Integrability in $N=2$ Gauge Theory: A Proof}, hep-th/9511052;

E.~Martinec, Phys. Lett. {\bf B367} (1996) 91;

R.~Donagi and E.~Witten, {\it Supersymmetric Yang-Mills Theory and 
Integrable Systems}, hep-th/9510101;

H.~Itoyama and A.~Morozov, {\it Integrability and Seiberg-Witten Theory:
Curves and Periods}, hep-th/951126; {\it Prepotential and 
the Seiberg-Witten Theory}, hep-th/9512161.


\end{thebibliography}
\end{document}